\documentstyle[12pt,aaspp4]{article}

\begin{document}

\title{The Narrow-Line Regions of LINERs as Resolved with the {\it Hubble 
Space Telescope}\footnote {Based on observations with the {\it Hubble Space
Telescope} which is operated by AURA, Inc., under NASA contract NAS
5-26555.}}

\author{Richard W. Pogge}
\affil{Department of Astronomy, The Ohio State University, 
140 W. 18th Ave., Columbus, OH 43210-1173}
\authoremail{pogge@astronomy.ohio-state.edu}

\author{Dan Maoz}
\affil{School of Physics \& Astronomy and Wise Observatory,
Tel-Aviv University, Tel-Aviv 69978, Israel}
\authoremail{dani@wise.tau.ac.il}

\author{Luis C. Ho}
\affil{Carnegie Observatories, 813 Santa Barbara St., Pasadena, 
CA 91101}
\authoremail{lho@ociw.edu}

\and

\author{Michael Eracleous}
\affil{Department of Astronomy and Astrophysics, The
Pennsylvania State University, 525 Davey Lab, University Park, PA 16802}
\authoremail{mce@astro.psu.edu}

\begin{abstract}

Low-ionization nuclear emission-line regions (LINERs) exist in the nuclei
of a large fraction of luminous galaxies, but their connection with the
active galactic nucleus (AGN) phenomenon has remained elusive.  We present
{\it Hubble Space Telescope} ({\it HST}) narrowband ([\ion{O}{3}]$\lambda
5007$ and H$\alpha$+[\ion{N}{2}]) emission-line images of the central
regions of 14 galaxies with LINER nuclei. This is the first such study of a
sizable sample of LINERs at {\it HST}\ resolution.  The compact,
$\sim$1\arcsec-scale, unresolved emission which dominates the line flux in
ground-based observations of these LINERs is mostly resolved in the {\it
HST}\ images. The bulk of the H$\alpha$ and [\ion{O}{3}] emission comes
from regions with sizes of tens to hundreds of parsecs.  The resolved
emission comes from a combination of knots, filaments, and diffuse gas
whose morphology differs from galaxy to galaxy. Most of the galaxies do not
show clear linear structures or ionization cones analogous to those often
seen in Seyfert galaxies.  An exception is NGC\,1052, the prototypical
LINER, in which we find a 3\arcsec-long ($\sim 250$ pc) biconical structure
that is oriented on the sky along the galaxy's radio jet axis.  M84 also
shows signs of possible biconical gas structures.  Seven of the galaxies
have been shown in previously published {\it HST}\ images to have a bright
compact ultraviolet (UV) nuclear source, while the other seven do not
display such a central UV source.  Our images show a dusty environment in
the nuclear region of all 14 galaxies, with clear indications of
obscuration of the nuclei in most of the ``UV-dark'' cases. The data thus
suggest that the line-emitting gas in most LINERs is photoionized by a
central source (which may be stellar, nonstellar, or a combination thereof)
but that this source is often hidden from direct view in the UV by dust in
the host galaxy.  We find no obvious differences between the morphologies
of the nine ``LINER 1.9s'' with detected weak broad H$\alpha$ wings in
their spectra and the morphologies of the other five objects. Likewise,
there is no clear distinction in morphology between objects whose UV
spectra are dominated by hot stars (e.g., NGC\,4569) and those that are
more AGN-like (e.g., NGC\,4579).

\end{abstract}


\section{Introduction}

Nuclear activity in galaxies, which finds its most dramatic expression in
quasars, also appears in systems with much lower luminosities.  Many
galactic nuclei exhibit broad H$\alpha$ emission lines which, while much
weaker, are nonetheless qualitatively similar to those observed in quasars
(Stauffer 1982; Keel 1983b; Filippenko \& Sargent 1985; Ho et al. 1997b). A
significant fraction of emission-line objects, which may be physically
related to AGNs, are galaxies containing low-ionization nuclear
emission-line regions (LINERs; Heckman 1980; see the reviews included in
Eracleous et al.  1996).  LINERs, present in over 30\% of all galaxies and
in 60\% of Sa--Sab spirals with $B\,\leq$ 12.5 mag (Ho, Filippenko, \&
Sargent 1997a), could thus represent the low-luminosity end of the AGN
phenomenon. In fact, about 15\%--25\% of LINERs have a broad component in
the H$\alpha$ line --- the ``type 1.9'' LINERs --- similar to the fraction
in Seyferts (Ho et al. 1997b).  Recently, Barth, Filippenko \& Moran
(1999a, b) have shown that some LINERs have weakly polarized broad emission
lines, analogous to the polarized broad lines from the ``hidden broad-line
region'' of some Seyfert 2 galaxies (e.g., Antonucci \& Miller 1985).
However, unlike Seyfert nuclei and QSOs, whose enormous luminosities and
rapid variability argue for a nonstellar energy source, the luminosities of
LINERs are sufficiently low that one cannot unambiguously associate them
with AGNs of higher luminosities.  For example, stellar energy sources are
plausible both on energetic and spectroscopic grounds (e.g., Terlevich \&
Melnick 1985; Filippenko \& Terlevich 1992; Shields 1992; Maoz et
al. 1998).  The potential role of LINERs in constituting the faint end of
the AGN luminosity function is important for understanding the nature of
AGNs, their evolution, and their contribution to the X-ray background.

To address some of the above issues, Maoz et al. (1995) obtained
ultraviolet (UV; 2300~\AA) images of an unbiased selection from a complete
sample of nearby galaxies with the {\it Hubble Space Telescope ({\it HST})}
Faint Object Camera (FOC).  They discovered that 6 out of 25 LINERs in the
sample contain unresolved ($<0\farcs1$, or $<1-2$~pc) nuclear UV emission
sources.  A similar result was found by Barth et al. (1998), using UV
images taken with the Wide-Field Planetary Camera 2 (WFPC2) on {\it HST}.
The extreme-UV emission from such sources may provide some or all of the
energy required to produce the nuclear emission lines by
photoionization. More specifically, Maoz et al. (1998) showed that in three
out of seven UV-bright LINERS, the extreme-UV flux, based on a reasonable
extrapolation from the UV, is sufficient to account for the observed
H$\alpha$\ flux.  In the other four objects, the extreme-UV flux is
deficient by a factor of a few, but these four objects have X-ray/UV flux
ratios 100 times larger than the previous three, which suggests that there
is much more flux in the extreme-UV than a simple extrapolation from the UV
would indicate. This suggestion is also supported by the spectral energy
distributions of LINERs and low-luminosity AGNs presented by Ho (1999). Any
mild foreground extinction would alleviate the deficit even further. It is
thus plausible to conclude that the line-emitting gas of UV-bright LINERs
is powered by photoionization.

The 6 UV-bright and 19 UV-dark LINERs studied by Maoz et al. (1995) are
otherwise similar in terms of spectral line ratios and overall
emission-line luminosities.  A nuclear UV source may therefore exist in all
LINERs, but may be obscured by dust in 75\% of the objects.  Alternatively,
Eracleous, Livio, \& Binette (1995) have suggested that the emission lines
are produced in response to a variable continuum that is in its ``off''
state with a 25\% duty cycle (due, perhaps, to sporadic tidal disruption
and accretion of individual stars by a central black hole).  Another
possibility is that the emission lines in UV-dark LINERs are produced in
shocked, rather than photoionized, gas (Koski \& Osterbrock 1976; Fosbury
et al. 1978; Heckman 1980; Dopita \& Sutherland 1995), thus accounting for
the absence of a central, point-like UV source.  Moreover, the UV-bright
LINERs are not necessarily AGNs, as the UV sources could be hot star
clusters. Indeed, UV spectroscopy with the {\it HST}\ has shown that, while
some LINERs may be AGNs (Ho et al. 1996; Barth et al.  1996), the UV
emission in other UV-bright LINERs is clearly dominated by massive stars
(Maoz et al. 1998). Interestingly though, there is not a clear
correspondence between the existence of a point-like nuclear UV source and
the detection of broad H$\alpha$\ wings in the spectrum, as is the case in
most Seyfert 1s.

An independent source of information comes from the X-ray band, where the
morphologies and spectra of LINERs suggest that some of them could harbor
low-luminosity AGNs.  Published and archival X-ray images of LINERs with
high angular resolution (5\arcsec--8\arcsec), taken with the {\it Einstein}
and {\it ROSAT} HRIs (e.g., Fabbiano, Kim, \& Trinchieri 1992; Koratkar et
al. 1995), show a wide variety of X-ray morphologies: point sources, with
or without a surrounding halo, and diffuse sources, which do not seem to be
related to the UV morphology.  The 0.5--10~keV spectra of LINERs obtained
with {\it ASCA} can generally be fitted by a linear combination of a
Raymond-Smith plasma model ($kT\approx 0.6-0.8$~keV) and an absorbed
(column densities in excess of $10^{21}$~cm$^{-2}$) hard component.  The
soft, thermal plasma emission is usually attributed to circumnuclear hot
gas.  In the case of LINER 1.9s, the hard component is well fitted by a
power law with photon indices $\Gamma\,\approx$ 1.7--2.0 (e.g.,
Serlemitsos, Ptak, \& Yaqoob 1996; Ptak et al. 1999; Awaki 1999; Terashima
1999; Ho et al. 1999), as seen in luminous Seyfert 1s (Nandra et al. 1997),
and the emission has a compact, spatially unresolved morphology within the
coarse angular resolution of {\it ASCA} (FWHM $\approx 3\arcmin$).  Where
higher resolution {\it ROSAT} HRI images are available, a central compact
core is seen in the soft X-rays as well.  These characteristics strengthen
the case that LINER 1.9s are genuine AGNs.  The situation for LINER 2s is
more complicated.  Terashima et al. (1999) have recently analyzed {\it
ASCA} observations of a small sample of LINER 2s, and they find that the
hard component, while consistent with a power law with $\Gamma\approx 2$,
can also be represented by a thermal bremsstrahlung model with a
temperature of several keV.  Moreover, the emission in the hard band is
seen to be extended on scales of several kpc, consistent with a population
of discrete sources such as low-mass X-ray binaries.  Terashima et al. also
show that, based on an extrapolation of their absorption-corrected X-ray
fluxes into the UV, there is perhaps insufficient power to drive the
luminosities of the optical emission lines.  These findings suggest that
either LINER 2s do not contain an AGN or that the AGN component, if
present, must be heavily obscured by matter with a column density much
greater than $10^{23}$~cm$^{-2}$.

Another important tool for studying AGNs, which we employ here, is
narrowband, emission-line imaging of the nuclear regions. Narrowband
imaging of Seyfert 1 and 2 nuclei has revealed, in some cases, striking
ionization cones emerging from the active nuclei and well aligned with the
axes of the radio jets (Haniff, Wilson, \& Ward 1988; Pogge 1989a; Wilson
\& Tsvetanov 1995).  This technique produces spectacular results when
combined with the angular resolution of {\it HST} (e.g., NGC\,5728, Wilson
et al. 1993).  The ionization structure of the narrow-line region gas, as
revealed by such studies, gives complementary information to that provided
by single-aperture spectra. Ground-based narrowband imaging of LINERs by
Keel (1983a) and by Pogge (1989b) has shown that they are distinct from
Seyferts in their circumnuclear emission, at least when probed on the same
($\sim 1$\arcsec) angular scale.  At these scales, some LINERs have faint
diffuse emission, as opposed to the linear structures in many Seyferts, and
the emission is usually dominated by a compact, marginally resolved nuclear
region.  Resolving the nuclear structures of LINERs can provide further
clues to their relation to AGNs. The small scales and faintness of these
structures relative to the bright host-galaxy background mean that the
capabilities of {\it HST} are needed for this task. To this end we have
carried out a study of the narrow-line regions of LINERs using narrowband
[\ion{O}{3}] $\lambda 5007$ and H$\alpha$+[\ion{N}{2}] WFPC2 images of 14
objects. The results of our study are the subject of this paper. In \S 2 we
describe the observations and the data reduction.  In \S 3 we present the
final images and measurements and we discuss them in \S4.  Finally, in \S5
we summarize the results and present our conclusions. Throughout this paper
we assume a Hubble constant of $H_0=75~{\rm km~s^{-1}~Mpc^{-1}}$.

\section{Observations and Reduction}

The galaxies we have chosen for narrowband imaging with {\it HST}\
represent the two classes of LINERs, UV-bright and UV-dark, that have been
identified in previous {\it HST}\ UV observations (Maoz et al. 1995; Barth
et al. 1998).  They were selected on the basis of their bright,
high-contrast H$\alpha$ lines, as determined from spectra and narrowband
images obtained from the ground. We have supplemented these data with
archival images of eight other galaxies, classified as LINERs by Ho et
al. (1997a), and observed with WFPC2 using the same narrowband filters. In
Table~1 we list the galaxies included in our collection, and we summarize
their basic properties.  We emphasize that these galaxies do not constitute
a statistically well-defined sample, but rather a random selection of
LINERs with relatively strong emission lines.  At any rate, this is the
first time that the spatial structure of the narrow-line region is studied
at {\it HST}\ resolution for a sizable number of such objects.

The galaxies were imaged with WFPC2, generally with the nucleus positioned
on the PC CCD, which has a scale of $0\farcs0455~{\rm pixel^{-1}}$. In the
case of NGC\,3031 the nucleus was positioned on the WF3 CCD, whose scale is
$0\farcs10~{\rm pixel^{-1}}$. Table 2 summarizes the observations of each
galaxy, gives the filters used, their corresponding exposure times, and the
observing programs under which the observations were carried out. All the
galaxies were observed through either the F656N or the F658N filters, in
order to sample the H$\alpha$+[\ion{N}{2}] complex at the proper redshift.
Images through the F502N filter, which covers the [\ion{O}{3}]$\lambda
5007$ line, exist for only five of these galaxies with sufficient
integration time to be useful.  Broad- and medium-band images of each
galaxy were also obtained, as detailed in Table 2, and used for continuum
subtraction and derivation of color maps. For NGC\,1052, whose narrowband
image was obtained from the {\it HST}\ archive, no broad-band images are
available.  However, the extended line emission in this object is strong
enough that is can be seen even without continuum subtraction. In fact,
this is the only object in which, after some additional image processing,
we find an unambiguous ionization cone, analogous to those seen in Seyferts
(see \S 3).

All of the WFPC2 images used in this study were processed by the standard
STScI OPUS pipeline (described by Biretta et al. 1996), and required only
minimal post-processing to combine multiple images, correct for saturated
pixels, and remove cosmic rays.  We present only the WFPC2 PC1 detector
images since the nuclei of the galaxies were centered on this CCD.  The
exception is NGC\,3031 (M81), for which the archival images had the nucleus
centered on the WF3 detector.  For the targets in our own observing program
(GO-6436), continuum images were acquired as two pairs of short and long
integrations.  If the object was particularly bright, an additional, 6~s
integration was also obtained.  The short-exposure images were used to
correct for saturated pixels in the long-exposure images.  Our narrowband
images were acquired as two or three long integrations, since saturation
was not expected to be a problem.

For pairs or triplets of images of the same integration time, we combined
the images using a statistical differencing technique implemented as an
{\tt XVista} command script (Pogge \& Martini 1999).  This technique is as
follows.  The difference image, formed by subtracting one image in a pair
from the other, consists primarily of positive and negative cosmic-ray
hits, as the galaxy, foreground stars, and background, all cancel to within
the noise.  All pixels within $\pm5\sigma$ of the mean residual background
level on the difference image are then set to zero (tagging them as
unaffected by cosmic rays), and a pair of cosmic-ray templates are derived
by separating the remaining positive and negative pixels.  These templates
are then subtracted from the original images, and the two cosmic-ray
subtracted images are added together to form the final galaxy image.  When
three images are available, all pair-wise combinations are used to generate
the templates.  In all cases, the statistical differencing method produced
superior cosmic-ray rejection compared to standard tasks (e.g., {\tt CRREJ}
in {\tt STSDAS}), and it is computationally much faster.

Archival data sets with pairs of images were processed in the same way. In
a few cases, however, only single integrations were available, and the
cosmic ray hits were removed manually, using the interactive {\tt TVZAP}
routine in {\tt XVista}.  When the archival images pairs had unequal
integration times (e.g., for NGC\,404, 500s and 1200s for the F656N
filter), we scaled the long integration to the shorter one and applied the
differencing method, followed by additional manual cleaning.  The resulting
image cleaning is not as thorough as with well-matched integration times,
but it still is better than the other algorithms we tried.

In all of our images, the mean intensity level of the background sky is
negligible (a few counts at most).  This was estimated by examining the
outskirts of one of the WF frames without much galaxy light in it, and
computing a modal sky level in reasonably clear regions.  The combined
on-band emission-line and off-band continuum images were converted to units
of flux density per pixel, based on the May 1997 updated photometry values
for each filter for the PC detector.

Continuum-subtracted emission-line images were created by subtracting the
associated continuum-band images.  In a few cases, it was clear that our
background estimate was in error (it left either positive or negative
fields of pixels), and so we refined the continuum estimation and iterated.
The final continuum-subtracted images were left in units of flux density
per pixel.  For several archival data sets, the continuum images had to be
registered and/or rotated to match the narrowband images.  This presented
no problem, and standard {\tt XVista} tools were used (the procedure is
analogous to the one described in Pogge 1992).

Color maps were generated for all six of our GO program images by
converting flux density per pixel into standard Johnson/Cousins magnitudes
using the transformations derived by Holtzman et al. (1995), and then
dividing the two images.  For NGC\,4192 and NGC\,4569, we had both F547M
and F791W image pairs from our own program as well as archival F555W and
F814W images, so we could verify the conversion between these bands and
estimates of the ($V-I$) colors.  We were careful to register the original
on-band and off-band images so that we could later directly compare our
emission-line and color maps.  We use these below to study the associations
between the emission-line regions and the patches of dust and star clusters
in the galaxies.  For the LINERs for which we have only archival images, we
could create ($V-I$) color maps for three galaxies (NGC\,3998, NGC\,4374,
and NGC\,4594).

For four galaxies with F547M images and no corresponding red broad-band
image (NGC\,3031, 4486, 4036, and 4258), we were able to map the
distribution of dust using an ``unsharp masking'' technique described by
Pogge \& Martini (2000).  In brief, an unsharp mask for an image was
created by smoothing the original F547M image with a model PSF image
computed using TinyTim (Krist \& Hook 1997).  The (Image\,{$\otimes$}\,PSF)
convolution was carried out in the Fourier domain using an {\tt XVista}
command script.  The original image was then divided by the smoothed image
to form a normalized residual image in which dusty features appear as
negative residuals, and emission or stars appear as positive residuals.
Using this technique on F547M images of galaxies for which we have $V-I$
color maps shows that the normalized unsharp residual images can retrieve
all of the dust structures seen in the color maps.

\section{Results}

\subsection{Images and Measurements}

Figures~1a--e show our reduced narrowband images and dust maps. For each
galaxy in Figure 1a--d we show the continuum-subtracted
H$\alpha$+[\ion{N}{2}] PC1 image on the left, and on the right show either
the $(V-I)$ color map, or an unsharp residual map of the F547M image if
there was no second broadband filter image.  In both the $V-I$ and the
F547M unsharp mask images darker shades denote the regions of dust
absorption.  Each panel of these figures shows a $10\arcsec\times10\arcsec$
segment of the image centered on the nucleus (oriented with North up and
East to the left), and with a scale bar in the lower left corner of each
emission-line image showing 100\,pc projected at the galaxy's distance (see
Table 1).  The contrast of the emission-line images is chosen to emphasize
the faint circumnuclear emission regions.  Figure 1e shows our images of
NGC\,3031 which, unlike the others, is on the WF3 detector.  Here we show
H$\alpha$+[\ion{N}{2}] emission on the left, and the unsharp residual map
of the F574M filter image on the right for the central 30\arcsec\ of this
galaxy.  The scale bar on the lower left shows 100\,pc at the distance of
NGC\,3031 (Table 1).  Figure 2 shows on the left the original F658N image
(i.e., without continuum subtraction) of NGC\,1052, with a normalized
unsharp residual map of the same on the right.  This residual map shows
emission as bright and absorption (presumably dust) as dark.  Each panel
shows the inner 15\arcsec\ of NGC\,1052, and the scale bar indicates
100\,pc.  The axis of the VLA radio jet (Wrobel 1984) is shown as a dashed
line.

Figures~3a and 3b show the continuum-subtracted [\ion{O}{3}] $\lambda5007$
images for the 5 galaxies for which these data are available.  In Figure
3a, we pair [\ion{O}{3}] $\lambda5007$ emission-line images with
``excitation maps'' of the H$\alpha$+[\ion{N}{2}]/[\ion{O}{3}] $\lambda
5007$ ratio for NGC\,4258, NGC\,4579, and NGC\,5005.  Although noisy, these
maps do not reveal any clear high-excitation knots with the exception of
NGC\,4258.  Here we see relatively-highly excited gas in a segment of the
braided jet that lies to the north of the nucleus in our images (Cecil,
Wilson, \& Tully 1992; Cecil, Wilson, \& DePree 1995; Cecil, Morse, \&
Veilleux 1995).  Figure 3b shows only the continuum-subtracted [\ion{O}{3}]
images for the remaining two galaxies, NGC\,4192 and NGC\,4569.  The
excitation maps constructed for these galaxies are extremely noisy due to
the low signal-to-noise ratio in the [\ion{O}{3}] images, and only total
fluxes in synthetic apertures can be measured with any confidence (see
Table 3).  Overall, the entire resolved emission-line regions of these
LINERs seem to be in a low-ionization state.

We have measured the integrated emission-line fluxes through various
apertures, separating the nuclear and circumnuclear contributions. These
measurements are summarized in Table~3.  Circular apertures were used,
except in the cases of NGC\,4192 and NGC\,5005, where rectangular apertures
were used to avoid strong dust lanes (in both) and \ion{H}{2} regions (in
NGC\,4192).  Since NGC\,4192 and NGC\,5005 have no discernible nuclei in
their narrowband images, the nuclear fluxes were estimated in apertures
centered using the brightness peaks in their F791W continuum images.
Table~3 gives ``band'' fluxes, without an attempt to convert to emission in
a particular line by correcting for the filter transmission of other lines
in the bandpasses (i.e., [\ion{O}{3}] $\lambda$4959 and
[\ion{N}{2}]~$\lambda\lambda$6548, 6583).

In the next section we describe the main features of the images of
individual galaxies.  We refer to an object as being ``UV-bright'' if
space-UV observations (generally the {\it HST}/FOC F220W images of Maoz et
al. 1995) have revealed a bright compact nuclear UV source in the
galaxy. We will base the optical spectral classification of these objects
on Ho et al. (1997a), and follow their terminology, where a ``LINER 2'' is
a LINER without detected broad H$\alpha$ wings, a ``LINER 1.9'' is a LINER
that does have such weak broad wings, and a ``transition object'' is one
whose optical narrow emission-line ratios are intermediate between those of
a LINER and an \ion{H}{2} nucleus. Data on these properties are also
summarized in Table 1.

\subsection{Individual Objects}

\subsection*{NGC\,404}

This is a UV-bright LINER 2, whose UV spectrum has a significant
contribution from massive stars (Maoz et al. 1998). In the {\it HST}\
emission-line images, much of the nuclear emission appears as a hollow
one-sided fan extending into filamentary wisps at distances of 5\arcsec\ or
more from the nucleus.  These wisps are reminiscent of gaseous structures
blown out by supernovae, which are expected, given that the spectrum of
this object is dominated by hot stars.  There is also one bright point
source 0\farcs16 north of the nucleus, possibly a planetary nebula or a
compact \ion{H}{2} region.  It is not obviously associated with a secondary
UV source seen in the FOC image of this galaxy.  The nuclear region is
dusty, but has a blue nucleus, suggesting the nucleus itself is unobscured.

The distance to this galaxy is controversial, as discussed in detail by
Wiklind \& Henkel (1990).  The distance we have adopted in Table~1 is
Tully's (1988) value of 2.4~pc (for $H_0=75~{\rm km~s^{-1}~Mpc^{-1}}$),
which was assigned based on NGC\,404's probable membership in his so-called
``14+12'' group. On the other hand, the CO observations of Wiklind \&
Henkel (1990) can only be reconciled with other observational data if the
distance is 10~Mpc.  The $I$-band (F814W) image shows the galaxy beginning
to be resolved into stars (the brightest giants are apparently visible). If
so, then our measurements favor the shorter distance.

\subsection*{NGC\,1052}

This galaxy is often considered to be the prototypical LINER, with weak
broad H$\alpha$ wings that give it a LINER 1.9 classification (Ho et
al. 1997a).  The broad wings have recently been shown by Barth et
al. (1999a) to be preferentially polarized relative to the narrow lines and
the continuum, suggesting the presence of a hidden broad-line region that
is seen in scattered light. Recent {\it HST}\ observations (Allen,
Koratkar, \& Dopita 1999; Gabel et al.  1999) show the UV-bright nucleus
has a UV--optical spectrum consisting of narrow lines on top of a
featureless continuum.  The archival H$\alpha$+[\ion{N}{2}] F658N WFPC2
image presented here shows a collimated, conical structure emerging from a
compact core. The high surface-brightness line emission is evident in the
original image even though we do not have a broad-band image to perform
continuum subtraction. The biconical nature of the structure is most
clearly brought out in the normalized unsharp residual map of the image
(Figure 2b), which reveals the rear (west) side of the cone.  Together with
M84 (see below), these are the only objects among the LINERs imaged which
show a clear indication of a Seyfert-like emission-line cone. The cone's
axis is at position angle 96\arcdeg\ with a full opening angle of about
70\arcdeg.  This corresponds roughly to the axis of the radio lobes
observed in this galaxy (Wrobel 1984), and is similar to the alignment
generally found in Seyferts. Our result thus adds another AGN
characteristic to this LINER. We also note that there are two faint knots
of emission straddling the nucleus, about 5\arcsec\ from it, along a
position angle of 81\arcdeg.

\subsection*{NGC\,3031 (M81)}

Detailed modeling of the narrow- and broad-line spectrum of this object (Ho
et al. 1996) clearly shows that the line-emitting gas has the
low-ionization state expected of LINERs, even if the measured
[\ion{O}{3}]/H$\beta$ ratio technically places it in the Seyfert class (Ho
et al.  1997a).  The UV spectrum (Ho et al. 1996; Maoz et al. 1998)
consists of broad, AGN-like, emission lines superposed on a featureless
continuum. Devereux, Ford, \& Jacoby (1997) have already presented the
H$\alpha$+[\ion{N}{2}] data shown here, and have also shown that the galaxy
possesses a UV-bright nucleus. The H$\alpha$+[\ion{N}{2}] emission comes
mostly from a bright compact source, surrounded by symmetric, disk-like,
diffuse emission (Figure 1e, left). The unsharp-residual processed image
(Figure 1e, right) shows a spiral-like dust lane extending $\sim$12\arcsec
north of the nucleus. The major axis of the disk is at a position angle of
18\arcdeg\ and has a minor-to-major axis ratio of 0.78. It extends up to
about 5\arcsec\ from the center, while faint, filamentary structure is
visible out to 8\arcsec.

\subsection*{NGC\,3718}

This is a UV-dark LINER 1.9. The emission in the {\it HST}\ narrowband
images is dominated by a strong point source, surrounded by some diffuse
circumnuclear H$\alpha$ emission.  The diffuse emission is brighter on one
side.  The $V-I$ image shows the nucleus is clearly very dusty and likely
obscured. This is not surprising given its very red optical spectrum (Ho et
al. 1995).

\subsection*{NGC\,3998}

This LINER 1.9 has been shown to be UV-bright by Fabbiano, Fassnacht,
\& Trinchieri (1994). Ultraviolet spectra from {\it HST}\ do not
exist, to date.  The H$\alpha$+[\ion{N}{2}] image shows a 100-pc
disk-like structure surrounding a compact nucleus. The major axis of
this disk is oriented along a position angle of 90\arcdeg\ with a
length of 3\arcsec, while the minor axis length is 2\arcsec.  The
$V-I$ map shows little indication of dust in the nuclear region.

\subsection*{NGC\,4036}

This LINER 1.9 is UV-dark, based on WFPC2 F218W images (Barth et
al. 1998). Its H$\alpha$+[\ion{N}{2}] image has a complex filamentary and
clumpy structure, with several ``tentacles'' extending up to 4\arcsec\
northeast of the nucleus along a position angle of 70\arcdeg. The nucleus
proper resembles an ellipse with a major axis of 0\farcs6 along a position
angle of 45\arcdeg.  The unsharp-masked F547M image reveals wisps of dust
in a disk-like configuration surrounding the nucleus on all scales
probed. This is one of the few LINERs in our sample whose emission-line
morphology can possibly be termed ``linear'' in some sense, but it seems
that this morphology is in the plane of the inclined dusty disk, rather
than perpendicular to it.

\subsection*{NGC\,4192 (M98)}

This object has been classified by Ho et al. (1997a) as a ``transition
object,'' one whose optical spectrum is intermediate between that of a
LINER and an \ion{H}{2} nucleus.  It appears dark in UV images. In the {\it
HST}\ emission-line images, the nucleus is resolved into knots spanning
0\farcs5 in an east-west direction. In the continuum images the nucleus
appears ``soft'', rather than having a sharp point source like NGC\,4569.
The $V-I$ map shows nuclear region is dusty and the nucleus probably
obscured. On larger scales, there is a ring of
\ion{H}{2} regions, partially obscured by dust, and already seen in
the ground-based images of Pogge (1989b).

\subsection*{NGC\,4258 (M106)}

This galaxy contains the famous masing disk (Watson \& Wallin 1994; Miyoshi
et al. 1995) whose Keplerian rotation provides some of the best evidence
for a massive black hole in a galactic nucleus. It has variably been
classified as a LINER or a Seyfert 1.9, and is another example of a
borderline case. Wilkes et al. (1995) and Barth et al. (1999c) have shown
that the spectrum in polarized light has emission lines that are broader
than the lines in the total flux spectrum. However, this is seen not only
in the Balmer lines but in most of the forbidden lines as well, with the
width of the lines in the polarized spectrum depending on the critical
density of the transition. The phenomenon is thus different from that of
the hidden broad-line regions revealed in polarized light in some Seyfert 2
galaxies.

A WFPC2 F218W image taken by Ho et al. (2000a) shows no conspicuous UV
nucleus.  We therefore aligned the brighter O/B star knots on the F218W
image with those in an archival F300W image of this galaxy.  We detect
2180~\AA\ flux from all the blue stars easily visible in the F300W and
F547M images (see Figure 4).  We then find that a ``nucleus'' {\it per se
is} visible in the F218W image, but it is weak and its contrast low
compared to its surroundings. Translation of the nuclear count rate to a UV
flux is not straightforward, because of the large time fluctuations in the
UV sensitivity of WFPC2, plus the proneness of the F218W filter to red
leaks when observing such obviously-red sources. The UV flux for this
nucleus, which we list in Table 1, accounts for neither effect and must
therefore be treated as uncertain. In any case, it is clear the flux is
quite low compared with the UV-bright objects in our sample.  It is
reasonable to treat this nucleus as intermediate, between UV-bright and
UV-dark.

The emission-line images show a compact core and a spiral feature emerging
to the north (extending up to 5\arcsec from the nucleus) which could be the
base of the helical emission-line jet seen on larger scales by Cecil,
Wilson, \& Tully (1992).  Thus, this may be considered another LINER with
collimated (or at least organized) narrow-line emission. Although there is
ample evidence for circumnuclear dust in the images, there is no dust that
obviously covers the nucleus in the unsharp-masked F547M image. This is
confirmed also in ``$U-V$'' image we have formed using the F330W and F547M
images.  Since the masing molecular gas disk is viewed nearly edge-on
(Miyoshi et al. 1995), with significant optical depth along the line of
sight to the nucleus, perhaps it is the dust in this disk itself that is
partially obscuring the nucleus in the UV, and thus making it appear so
weak.

\subsection*{NGC\,4374 (M84)}

The LINER 2 nucleus of this galaxy is UV-dark, based on FOC F220W imaging
by Zirbel \& Baum (1998).  M84 has a nonthermal, flat-spectrum radio core
and compact X-ray emission (see discussion in Ho 1999), and its nucleus has
recently been found to contain a massive compact dark object, presumably a
supermassive black hole (Bower et al. 1998).  The H$\alpha$+[\ion{N}{2}]
data have been previously presented by Bower et al. (1997).  The images
show an inclined gas disk surrounding the nucleus. Our $V-I$ map clearly
shows that the nucleus is covered by a thick dust lane. Bower et al. (1997)
also argued for the possible presence of an ionization cone that is roughly
aligned with the radio structure in this object (Birkinshaw \& Davies
1985), but we find the case for such a cone is not clear.  At the very
least, it is not an obvious morphological structure in the extended
H$\alpha$ emission-line gas (Figure 1c, top left panel). This structure
takes the form of filaments that extend roughly east-west and north-south,
along position angles 85\arcdeg\ and 0\arcdeg. The east-west complex
extends 5\arcsec\ east and 3\arcsec\ west of the nucleus, while the
north-south complex extents 2\arcsec\ north and south of the nucleus.

\subsection*{NGC\,4486 (M87)}

This LINER 2, a giant elliptical galaxy in the Virgo cluster, is well known
for its collimated jet seen at radio, optical, and UV wavelengths.  Both
the jet and the dynamical evidence for a supermassive black hole (Sargent
et al.  1978; Harms et al. 1994; Macchetto et al. 1997) testify to the
existence of an AGN. The nucleus is UV-bright (Boksenberg et al. 1992; Maoz
et al. 1996).  Recent UV spectroscopy of the nucleus with {\it HST}/FOS
(Sankrit, Sembach, \& Canizares 1999) and {\it HST}/STIS (Ho et al. 2000b)
reveals emission lines of width $\sim 3000$ km s$^{-1}$ on top of a
featureless continuum. The H$\alpha$+[\ion{N}{2}] image was previously
published by Ford et al. (1994). It shows a compact disk with a major axis
of length 0\farcs77 along position angle 0\arcdeg, and a minor axis of
length 0\farcs59. The disk is surrounded by wispy filaments extending in
various directions up to 10\arcsec\ from the nucleus.  It is noteworthy
that the optical jet, which is conspicuous in the raw data (and also
visible in the unsharp residual map in Figure 1c), disappears completely in
the continuum-subtracted image, indicating very little line emission from
the jet itself. The unsharp residual map also shows very little evidence of
nuclear dust.

\subsection*{NGC\,4569 (M90)}

This galaxy has a bright, point-like nucleus at optical and UV bands, with a
LINER 2 optical spectrum. Maoz et al. (1998) have shown that the UV
spectrum is dominated by massive stars.  The new {\it HST}\ images show an
unresolved nucleus in both continuum and emission lines.  The nucleus
dominates the emission. On larger scales, there is a disk or
spiral-arm-like structure in the H$\alpha$ image, extending up to 2\arcsec\
from the nucleus in the north-south direction (position angle 4\arcdeg).
Similar structures are seen in [\ion{O}{3}] although they are not as well
defined. The $V-I$ map shows that, while the circumnuclear region is dusty,
the nucleus itself is apparently unobscured by dust.

\subsection*{NGC\,4579 (M58)}

This is a LINER 1.9 galaxy with many AGN characteristics (Filippenko
\& Sargent 1985; Barth et al. 1996; Ho et al. 1997b; Maoz et al. 1998;
Terashima et al.  1998). The H$\alpha$ emission is dominated by a
nuclear point source, but is surrounded by complex clumpy and
filamentary emission.  The overall complex has an elliptical shape
with a major axis of length 2\arcsec\ along position angle 120\arcdeg\
and a minor axis of length 1\arcsec.  The filamentary emission may be
likened to a shell or a ring (perhaps part of a disk) with a dark lane
going across it. A similar structure is seen in [\ion{O}{3}], although
the signal-to-noise ratio is lower. The $V-I$ image shows that, while
the filaments are associated with circumnuclear dust, the nucleus
appears to be unobscured.

\subsection*{NGC\,4594 (M104)}

The ``Sombrero'' galaxy has a LINER 2 nucleus which may be, like NGC\,4258,
borderline between UV-bright and UV-dark.  Crane et al. (1993) have shown
that the nucleus appears unresolved and isolated in {\it HST}\ images at
3400~\AA. However, the {\it HST}/FOS UV spectrum of this galaxy, analyzed
by Nicholson et al. (1998) and Maoz et al. (1998), shows shortward of
3200~\AA\ a red continuum falling with decreasing wavelength, and becoming
dominated by scattered light within the spectrograph below around
2500\AA. In Table~1 we quote the flux density measured by Maoz et
al. (1998) from this spectrum, but because of the scattered light
contamination and the lack of a UV image, we regard the quoted flux density
as an upper limit to the true value. Due to the low signal-to-noise ratio
of the UV spectrum, the nature of the UV light source (stars or AGN) is
ambiguous.  Fabbiano
\& Juda (1997) observed this galaxy with the {\it ROSAT}/HRI and
detected a point-like soft X-ray source coincident with the nucleus but
noted that the source could be highly absorbed.  The
H$\alpha$+[\ion{N}{2}] image shows ``S''-shaped wisps emerging from a
bright, compact, possibly disky H$\alpha$ core. The two wisps extent
up to 4\arcsec\ east and west and up to 1\arcsec\ south of the
nucleus.  The $V-I$ image shows that the dust generally follows the
H$\alpha$ morphology, but with the nucleus behind a dust lane.

\subsection*{NGC\,5005}

This is a LINER 1.9, which is dark in UV images. In the new {\it HST}\
images, the line emission is distributed in a number of compact clumps
within 1\arcsec of the nucleus. These are surrounded by fan-shaped
filaments and diffuse emission extending up to 3\arcsec\ southeast of the
nucleus. The emission-line and $V-I$ images both show clearly that the
nucleus is obscured. In an attempt to identify whether the emission-line
clumps are associated with individual stars or star clusters, we have tried
to align the H$\alpha$+[\ion{N}{2}] image with the FOC 2200\AA\ image of
Maoz et al. (1996). We find no unique registration that will align all the
major {\ion{H}{2}} regions and the UV knots in a region 10\arcsec\ south of
the nucleus, and no registration that can align the nuclear UV and
H$\alpha$ knots. It thus appears that here, as in the other galaxies, the
line-emitting gas is dusty, causing the UV and H$\alpha$ emission to be
mutually exclusive. As a consequence, we cannot answer conclusively the
question of whether, in this galaxy, there is direct evidence for the
excitation of the emission-line gas by hot stars.

\section{Discussion}

With the information given above, we are in a position to address some of
the following questions.

\begin{enumerate}

\item Do any of the LINERs, when observed at {\it HST} resolution,
      show ionization cones or linear structures analogous to those seen in
      Seyferts? If so, what are their general characteristics (e.g.,
      opening angles, linear extent, excitation level)? Ionization cones
      (or lobes) are probably the best evidence for obscuration of the
      nucleus by a toroidal structure, which would account for the absence
      of a nuclear UV source in UV-dark LINERs.

\item Is there a difference in the morphology of the ionized gas in the 
      circumnuclear regions of UV-bright and UV-dark LINERs? Differences in
      morphology can afford direct tests of competing scenarios, as
      follows:

\begin{enumerate}
\item {\it Obscuration:} The nuclear UV source could be hidden by a toroidal
      structure, as detailed above, or with patchy foreground obscuration
      by circumnuclear dust (e.g. van Dokkum \& Franx 1996), not
      necessarily associated with the nucleus itself.

\item {\it An ionizing continuum source temporarily in its ``off'' state}: The
      duty-cycle hypothesis of Eracleous et al. (1995) predicts a spatial
      gap between the nucleus and the ionization front in the
      [\ion{O}{3}]-emitting region because of rapid recombination of the
      O$^{+2}$ ion.  In contrast, the long recombination time scale of the
      ionized zone implies that its corresponding gap should be
      unobservable across the narrow-line region.  The recurrence time of
      active phases of the nuclear source in this scenario is of order a
      century. In view of the distances of these galaxies the implied
      angular size of a typical [\ion{O}{3}] ring would be around
      0.$^{\prime\prime}$6, well within the resolution of these {\it HST}
      images.

\item {\it Shock excitation of the emission-line gas}: This could manifest 
      itself as filamentary and bow-shaped structures indicative of shock
      fronts.  The emission-line images can be particularly informative at
      scales of a few arcseconds where the high angular resolution of the 
      {\it HST} and the often-seen clumpiness of line-emitting gas can reveal
      faint line-emitting structures that are undetectable from the ground.
\end{enumerate}
\end{enumerate}

First, we find that only one of the LINER nuclei observed, NGC\,1052, shows
an unambiguous ionization cone of the kind often seen in Seyfert
galaxies. M84 may also exhibit a biconical structure, but the evidence in
that object is less clear.  Two other galaxies, NGC\,4036 and NGC\,4258,
have structures that could plausibly be termed ``linear.''  None of the
remaining 10 LINERs show this kind of morphology.  Our attempt to find a
link between LINERs and AGNs through this avenue has therefore given a
positive result in only one, or at most four, cases. In NGC\,1052, which
already has various known AGN features, the cones are indeed aligned with
the radio structure, as in Seyferts.  Similarly, the possible biconical gas
structure in M84, if real, would be roughly aligned with the axis of its
radio jets (Birkinshaw \& Davies 1985).  Obviously, in the other objects we
cannot search for alignment of the complex emission line structures with
radio structures.  Nonetheless, it will be interesting to see in the future
whether or not elongated radio structures are common in LINERs.

Second, there is no clear difference in emission-line morphology between
the UV-dark and UV-bright LINERs, but rather, there is a large variety from
object to object. On the other hand, there is clear evidence for
obscuration of the nucleus by clumps and lanes of dust in all of the
clearly UV-dark objects, but not in the UV-bright ones. We conclude that
foreground obscuration by nuclear dust is the cause of the non-detection of
a central UV point source in these LINERs, if such a source is present. In
the one possible exception, NGC\,4258, the detected but weak central UV
source may be attenuated by dust mixed with the molecular gas in the masing
disk that is known to exist on the line of sight to the nucleus.  Although
our sample is small and statistically incomplete, one may speculate that
this is the reason that 75\% of LINERs are UV-dark (Maoz et al. 1995; Barth
et al. 1998) --- that is, that all LINERs are photoionized by a central UV
source, whose nature is of yet unknown, but that this source is obscured by
circumnuclear dust in 75\% of the cases.

In the same vein, we have found no evidence for obscuration by toroidal
structures on smaller scales (which would produce the ionization cones we
have generally failed to find), nor signs of a central source with ``gaps''
in the gas morphology hypothesized by Eracleous et al. (1995) in their
duty-cycle picture.  Nor do we find clear signs of outflows and shock-like
morphologies, although there are hints of structures that may turn out to
be related to such phenomena, if studied with deeper images at higher
resolution.  If, as the above results suggest, all LINERs have a central UV
source with a photon flux of the right order of magnitude to power the
observed emission line spectrum, then shocks are not needed to explain the
excitation of the emission-line gas.

Our sample contains similar numbers of so-called LINER 1.9s, i.e., LINERs
with weak broad H$\alpha$ emission, and LINER 2s, in which such broad lines
have not been detected.  The relative numbers of these two types among the
LINER population are similar to the relative numbers of Seyfert~1 and
Seyfert~2 galaxies (Ho et al. 1997b), and this may be another clue to a
relation between LINERs and higher-luminosity AGNs. We find, however, no
obvious differences in the emission line morphologies of the two LINER
types.  This is contrary to Seyferts, where the line morphologies of
Seyfert 1s are more compact (Pogge 1989b; Schmitt \& Kinney 1996),
suggestive of a geometry in which the central engine and broad-line region
are viewed unobscured along the axis of an obscuring torus.  A caveat to
this point is that the above study has compared Seyfert 1s and 2s, rather
than 1.9s and 2s, and this distinction may be important.

One can imagine a number of physical reasons for the differences in
the morphologies of LINERs and Seyferts. LINERs may, as a general
rule, lack the toroidal collimating structures postulated in
Seyferts. Alternatively, they may generally lack the relativistic jets
that are often coaligned with extended emission structures in
Seyferts. The jet/emission-line region alignment in Seyferts is
thought to arise because both jets and ionizing radiation are
collimated by related structures, or because the jet opens a path
through the interstellar medium for ionizing photons to follow, or
because the jet itself excites the line emission.  Among our sample,
this explanation cannot apply to M87, which has a conspicuous jet, yet
no linear emission-line structure, either coincident with the jet or
elsewhere.

Another possible explanation for the difference between LINERs and Seyferts
is a deficit of circumnuclear gas or of ionizing photons in LINERs on
the larger scales where linear structures appear in Seyferts.  However, all
Seyferts with extended narrow-line regions that have been imaged at {\it
HST}\ resolution to date show that the collimated linear structures and
cones persist all the way to the smallest angular scales probed (NGC\,1068:
Axon et al. 1998; NGC\,4151: Evans et al. 1993; NGC\,5252: Tsvetanov et
al. 1996), and this is also what we have found in the biconical emission of
the LINER NGC\,1052. On the other hand, one might argue that these objects
were preselected to have the narrowest and brightest extended narrow-line
regions, and do not represent the Seyfert population as a whole. Finally,
we note that the absence of linear emission-line structures in LINERs
do {\it not} preclude them from being AGNs. Indeed, many of the LINERs in our
sample have radio jets and/or broad-line regions, features that are
considered characteristic of nuclear activity in more powerful objects.
While linear emission-line features are found in many powerful AGNs, they
are by no means a defining characteristic of the class.

A further point that has interesting physical and practical implications is
that, when imaged at {\it HST}\ resolution, LINERs do not reveal simple
disk-like gas structures, but rather more complex geometries. This implies
that the kinematics of the circumnuclear gas are also likely to be quite
complicated and could lead astray the interpretation of kinematic
measurements aimed at determining the central black hole masses. Of special
interest is the morphology of the line-emitting gas in the innermost
regions close to the nucleus.  In most of the cases in our sample, there is
no indication of a small-scale disk, even if such a disk exists on larger
scales.  The kinematics of the gas at small radii, therefore, is unlikely
to be governed predominantly by rotation.  Indeed, recent {\it HST}\
spectroscopy of several galactic nuclei shows that the ionized gas has
velocity dispersions that are large even in the innermost regions, as
opposed to the circular velocity field expected from a cold gas disk.  For
example, the {\it HST FOS} emission-line spectra of the nuclear ionized gas
disk in M87 (Harms et al. 1994; Macchetto et al. 1997) show line widths of
$\sigma\approx 500~{\rm km~s^{-1}}$ at projected radii of 0\farcs2 to
0\farcs6 where the rotational velocity is $500-600{\rm km~s^{-1}}$ (see
Figure 5 of Macchetto et al. 1997).  A similar trend is seen in NGC\,4261
(Ferrarese, Ford, \& Jaffe 1996): at a deprojected distance of $\sim
0\farcs2$ from the nucleus, the gas disk shows $v/\sigma\approx 1$.
Finally, the nuclear ionized gas disk of M84 observed by Bower et
al. (1998) also displays large nonrotational motions near the center.

It is puzzling how gas in such a disturbed kinematic state within such a
small ($\sim 10^3$ pc$^3$) volume can avoid settling into a cool,
rotationally-dominated disk. The clumpy gas filaments will collide with
each other at supersonic velocities of order 100--200 km s$^{-1}$ on a
dynamical time scale, which at a distance of 5 pc from a $10^8 M_{\odot}$
central mass is $10^5$ yr. This is much shorter than the expected lifetime
of the AGN or the nuclear starburst, but much longer than the cooling time,
which, for free-free emission, is of order 100 yr for gas with a density
$10^5~{\rm cm^{-3}}$ that has been heated to $\sim 10^6$~K by collisions.

\section{Summary}

We have presented narrowband ([\ion{O}{3}]$\lambda 5007$ and
H$\alpha$+[\ion{N}{2}]) emission-line images of 14 galaxies with LINER
nuclei.  Most of these data have not been previously published, and this is
the first time that the narrow-line regions of a significant number of
LINERs are studied at {\it HST}\ resolution. The objects in our sample
include representatives of the various subclasses of LINERs that have
emerged in recent years: ``type 1.9'' and ``type 2,'' UV-bright and
UV-dark, objects with starburst-dominated or AGN-dominated UV spectra.

Our main observational findings are as follows.

\begin{enumerate}

\item The narrow-line regions of nearby LINERs are resolved by {\it HST}, with
      much of the line emission coming from regions with sizes of
      10--100~pc.

\item In general, the emission-line morphology is complex and disordered,
      with varying contributions from a compact core, a disk, clumps, and
      filaments.  We find no obvious distinctions in morphologies among the
      various LINER subclasses.

\item In only one object, NGC\,1052, possibly two if we include M84, have we
      found clear evidence for an ionization cone analogous to those seen
      in Seyfert galaxies. The ionization cone of NGC\,1052 is well-aligned
      with its radio structure.  Two or three other objects have
      morphologies that can perhaps be termed ``linear.''

\item Obscuration of the nucleus by circumnuclear clumps of dust is fairly
      ubiquitous in the UV-dark LINERs but absent in the UV-bright ones.

\end{enumerate}

These findings lead us to the following conclusions. First, the data are
consistent with a picture in which most or all LINERs are objects that are
photoionized by a central UV source, even when the central source is not
visible directly. As discussed in \S1, Maoz et al. (1998) showed that in
UV-bright LINERS, the extreme-UV flux, based on a reasonable extrapolation
from the UV, is of the right magnitude to account for the observed
H$\alpha$\ in a photoionization scenario.  Any mild foreground extinction,
which appears to be common based on the images presented here, would only
strengthen this conclusion.  Hence, the line emission UV-bright LINERs is
likely to be powered by photoionization.  Our results suggest that the UV
visibility of the nucleus is determined simply by the circumnuclear dust
morphology along our line of sight. This suggestion is reinforced by the
the anti-correlation between UV brightness on the one hand, and galaxy
inclination and Balmer decrement on the other, found by Barth et
al. (1998).  A similar inclination effect has been seen in Seyfert galaxies
at visible (Keel 1980) and X-ray wavelengths (Lawrence \& Elvis 1982).
Thus, the fact that the majority of the LINERs in optically-selected
samples are UV dark (Maoz et al. 1995; Barth et al. 1998) does not
necessarily imply that these objects are excited by processes other than
photoionization (e.g., shocks; Dopita \& Sutherland 1995) or that they are
in an ``off'' state (Eracleous et al. 1995). There is also no
correspondence between UV darkness and the absence of broad lines in
LINERs, which one might expect in a duty-cycle scenario when the continuum
source is turned off. Moreover, the UV spectra of the nuclei of individual
LINERs have so far failed to reveal the emission-line signatures predicted
by shock models.  This result argues against these alternative explanations
for the UV-bright/dark dichotomy. If, as our results suggest, UV-dark
LINERs appear as such only because of foreground extinction, then it is
plausible to conclude that all LINERs harbor a source of ionizing
radiation, and hence that their line-emitting gas is powered by
photoionization.

In passing, we note that in the non-elliptical galaxies in our sample, the
circumnuclear dust, though patchy and sometimes chaotic in appearance,
generally lies in a preferred plane.  The position angle of this plane
coincides remarkably closely to the direction of the major axis of the
large-scale galactic disk (data compiled in Ho et al. 1997a).  This
explains why the UV visibility of the nuclei correlates with the
inclinations of the host galaxies (Barth et al. 1998) despite the fact that
the obscuration, as seen in our images, actually occurs on much smaller
scales.

Second, whatever the nature of the central source in a LINER, be it an
accretion flow onto a black hole, a compact star cluster, or a combination
of the two, it is not generally revealed by the narrowband images we have
obtained.  The one LINER that shows a clear Seyfert-like ionization cone,
NGC\,1052, does have additional AGN characteristics: weak broad wings in its
H$\alpha$ emission profile (Ho et al. 1997b), a hidden broad-line region
(Barth et al. 1999a), a radio jet and compact, flat-spectrum core (Wrobel
1984), and a nonthermal hard X-ray spectrum (Guainazzi \& Antonelli 1999;
Weaver et al.  1999). On the other hand, many of the other LINERs which
have AGN features, such as NGC\,4579, M81, and M87, show no evidence for
ionization cones in our images.

Finally, we have pointed out that the complex gas morphologies revealed by
our images suggest caution in interpreting the gas kinematics in the
innermost regions of these objects, for example, in searches for, and mass
measurements of, central black holes.

\acknowledgements
This work was supported by grant GO-06436.01-95A from the Space Telescope
Science Institute, which is operated by AURA, Inc., under NASA contract NAS
5-26555.  Undergraduate research assistant S. Benfer (Ohio Wesleyan) helped
with the initial reductions of our GO imaging data.  D.M. acknowledges
support by a grant from the Israel Science Foundation.  L.C.H. is grateful
to Sandra Faber for bringing to his attention the issue concerning the
kinematics of the compact narrow-line regions in LINERs that we discussed
at the end of \S3.2.

%
%

\clearpage

\noindent{{\bf Figure 1a-d}: Narrowband PC1 images and dust maps. For each galaxy we
show the continuum-subtracted H$\alpha$+[\ion{N}{2}] image on the left, and
on the right either the $(V-I)$ color map or the unsharp mask of the F547M
frame if no $I$-band image is available.  Darker shades denote regions of
dust absorption.  Each panel shows a 10\arcsec$\times$10\arcsec\ segment of
the image centered on the nucleus (oriented with North up, East to the
left), and with a scale bar in the lower left corner of each emission-line
images showing 100\,pc projected at the galaxy's distance (see Table 1).
The contrast of the emission-line images is chosen to emphasize the faint
circumnuclear emission regions.}

\noindent{{\bf Figure 1e}: Narrowband WF3 images of the central 30\arcsec\ of
NGC\,3031, shown like the others in Figure 1a-d.  H$\alpha$+[\ion{N}{2}]
emission is on the left, and the unsharp residual map is on the right.  The
scale bar in the lower left shows 100\,pc at the distance of NGC\,3031
(Table 1).}

\noindent{{\bf Figure 2}: Narrowband PC1 images of the central 15\arcsec\ of
NGC\,1052.  Left panel: F658N image (without continuum subtraction); Right
panel: normalized unsharp residual map of the same.  The residual map shows
emission as bright and absorption (presumably dust) as dark.  Each panel
shows of NGC\,1052, and the scale bar indicates 100\,pc.  The axis of the
VLA radio jet (Wrobel 1984) is shown with the dashed line.}

\noindent{{\bf Figure 3a}: Continuum-subtracted PC1 [\ion{O}{3}]$\lambda5007$
emission-line images (left) of NGC\,4258, NGC\,4579, and NGC\,5005, shown
alongside of ``excitation maps'' of the H$\alpha$+[\ion{N}{2}]/[\ion{O}{3}]
$\lambda 5007$ ratio (right).  The scaling and orientation follow that in
Figures 1a-d.}

\noindent{{\bf Figure 3b}: Continuum-subtracted PC1 [\ion{O}{3}]$\lambda5007$ images
of NGC\,4192 (left) and NGC\,4569 (right).  For these galaxies the
``excitation maps'' are extremely noisy and contain no useful
information. The scaling and orientation are as in Figure 3a.}

\noindent{{\bf Figure 4}: Montage of PC1 images of NGC\,4258 showing (left-to-right,
top-to-bottom) F547M, [\ion{O}{3}]$\lambda5007$ emission, F300W, and F218W.
All maps show the central 10\arcsec of the galaxy centered on the active
nucleus.  The scaling and orientation are as in Figures 1--3.}

%
%


\begin{references}
\reference{} Allen, M. G., Koratkar, A. P., \& Dopita, M. A. 1999, \baas, 194,
             4901
\reference{} Antonucci, R. R. J., \& Miller, J. S. 1985, \apj, 297, 621
\reference{} Awaki, H. 1999, Advances in Space Research, 23 (5-6), 837
\reference{} Axon, D. J., Marconi, A., Capetti, A., Maccetto, D. F., 
             Schreier, E., \& Robinson, A. 1998, \aap, 496, L75
\reference{} Barth, A. J., Filippenko, A. V., \& Moran, E. C. 1999a, \apjl, 
             515, L61
\reference{} Barth, A. J., Filippenko, A. V., \& Moran, E. C. 1999b, \apj, 
             in press (astro-ph/9905290)
\reference{} Barth, A. J., Ho, L. C., Filippenko, A. V., \& Sargent, W. L. W. 
             1998, \apj, 496, 133
\reference{} Barth, A. J., Reichert, G. A., Filippenko, A. V., Ho, L. C.,
             Shields, J. C., Mushotzsky, R. F., \& Puchnarewicz, E. M. 1996,
             \aj, 112, 1829
\reference{} Barth, A. J., Tran, H., Brotherton, M. S., Filippenko, A. V., Ho, 
             L. C., van Breugel, W., Antonucci, R., \& Goodrich, R. W.
             1999c, \apj, in press (astro-ph/9907269)
\reference{} Biretta, J. A., et al. 1996, WFPC2 Instrument Handbook 
             (Baltimore: STScI)
\reference{} Birkinshaw, M. \& Davies, R. L. 1985, \apj, 291, 32
\reference{} Boksenberg, A., et al. 1992, \aap, 261, 393
\reference{} Bower, G. A., Heckman, T. M., Wilson , A. S., \& Richstone, D. O.
             1997, \apjl, 483, L33
\reference{} Bower, G. A., et al. 1998, \apjl, 492, L111
\reference{} Cecil, G., Wilson, A. S., Tully, R. B. 1992, \apj, 390, 365
\reference{} Cecil, G., Wilson, A. S., \& DePree C. 1995, \apj, 440, 181
\reference{} Cecil, G., Morse, J. A., \& Veilleux, S. 1995, \apj, 452, 613
\reference{} Crane, P., et al. 1993, \aj, 106, 1371
\reference{} Devereux, N., Ford, H., \& Jacoby, G. 1997, \apjl, 481, L71
\reference{} Dickey, J. M. \& Lockman, F. J. \araa, 28, 215
\reference{} Dopita, M. A. \& Sutherland, R. S. 1995, \apj, 455, 468
\reference{} Eracleous, M., Livio, M., Binette, L. 1995, \apjl, 445, L1
\reference{} Eracleous, M., Koratkar, A., Leitherer, C.,  \& Ho,  L. 1996, 
             The Physics of LINERs in View of Recent Observations (San
             Francisco: ASP)
\reference{} Evans, I. N., Tsvetanov, Z., Kriss, G. A., Ford, H. C., 
             Caganoff, S., \& Koratkar, A. P. 1993, \apj, 417, 82
\reference{} Fabbiano, G., Fassnacht, C., \& Trinchieri, G. 1994, \apj, 434, 67
\reference{} Fabbiano, G. \& Juda, J. Z. 1997, \apj, 476, 666
\reference{} Fabbiano, G., Kim, D.-W., \& Trinchieri, G. 1992, \apjs, 80, 531
\reference{} Ferrarese, L, Ford, H. C., \& Jaffe, W. 1996, \apj, 470, 444
\reference{} Filippenko, A. V. \& Sargent, W. L. W. 1985, \apjs, 57, 503
\reference{} Filippenko, A. V. \& Terlevich, R. 1992, \apjl, 397, L79
\reference{} Fosbury, R. A. E., Melbold, U., Goss, W. M., \& Dopita, M. A. 
             1978, \mnras, 183, 549
\reference{} Freedman, W. L., et al. 1994, \apj, 427, 628
\reference{} Ford, H. C., et al. 1994, \apjl, 435, L27
\reference{} Gabel, J. R., Bruhweiler, F. C., Crenshaw, D. M., Kraemer, S. B., 
             \& Miskey, C. L. 1999, \baas, 194, 4902
\reference{} Guainazzi, M. \& Antonelli, L. A. 1999, \mnras, 304, L15
\reference{} Haniff, C. A., Wilson, A. S., \& Ward, M. J. 1988, \apj, 334, 104
\reference{} Harms, R. J., et al. 1994, \apjl, 435, L35
\reference{} Heckman, T. M. 1980, \aap, 87, 152
\reference{} Herrnstein, J. R., et al. 1999, \nat, in press (astro-ph/9907013)
\reference{} Ho, L. C. 1999, \apj, 516, 672
\reference{} Ho, L. C., et al. 2000b, in preparation
\reference{} Ho, L. C., Filippenko, A. V., \& Sargent, W. L. W. 1995, 
             \apjs, 98, 477
\reference{} Ho, L. C., Filippenko, A. V., \& Sargent, W. L. W. 1996,
             \apj, 462, 183
\reference{} Ho, L. C., Filippenko, A. V., \& Sargent, W. L. W. 1997a, 
             \apjs, 112, 315
\reference{} Ho, L. C., Filippenko, A. V., Sargent, W. L. W., \& Peng, C. Y. 
             1997b, \apjs, 112, 391
\reference{} Ho, L. C., Filippenko, A. V., \& Sargent, W. L. W.  2000a, 
             in preparation
\reference{} Ho, L. C., Ptak, A., Terashima, Y., Kunieda, H., Serlemitsos, 
             P. J., Yaqoob, T., \& Koratkar, A. P. 1999, \apj, in press 
	     (astro-ph/9905013)
\reference{} Holtzman, J. A., et al. 1995, \pasp, 107, 1065
\reference{} Keel, W. C. 1980, \aj, 85, 198
\reference{} Keel, W. C. 1983a, \apj, 268, 632
\reference{} Keel, W. C. 1983b, \apj, 269, 466
\reference{} Koratkar, A. P., Deustua, S., Heckman, T. M., Filippenko, A. V., 
             Ho, L. C. \& Rao, M. 1995, \apj, 440, 132
\reference{} Koski, A. T. \& Osterbrock, D. E. 1976, \apjl, 203, L49
\reference{} Krist, J. \& Hook, R. 1997, The Tiny Tim User's Guide,
             Version 4.4 (Baltimore: STScI)
\reference{} Lawrence. A. \& Elvis, M. 1982, \apj, 256, 410
\reference{} Macchetto, F., Marconi, A., Axon, D. J., Capetti, A., Sparks, 
             W. B., \& Crane, P. 1997, \apj, 489, 579
\reference{} Maoz, D., Filippenko, A. V., Ho, L. C., Rix, H.-W., Bahcall, 
             J. N., Schneider, D. P., \& Macchetto, F. D. 1995, \apj, 440,
             91
\reference{} Maoz, D., Filippenko, A. V., Ho, L. C., Macchetto, F. D.,
             Rix, H.-W., \& Schneider, D. P. 1996, \apjs, 107, 215
\reference{} Maoz, D., Koratkar, A. P., Shields, J. C., Ho, L. C., Filippenko, 
             A. V., \& Sternberg, A. 1998, \aj, 116, 55
\reference{} Miyoshi, M., Moran, J., Herrnstein, J., Greenhill, L., Nakai, N., 
             Diamond, P., \& Inoue, M. 1995, \nat, 373, 127
\reference{} Nandra, K., George, I. M., Mushotzky, R. F., Turner, T. J.,
             \& Yaqoob, T. 1997, \apj, 477, 602
\reference{} Nicholson, K. L., Reichert, G. A., Mason , K. O., Puchnarewicz, 
             E. M., Ho, L. C., Shields, J. C., \& Filippenko, A. V. 1998,
             \mnras, 300, 893
\reference{} Pogge, R. W. 1989a, \apj, 345, 730
\reference{} Pogge, R. W. 1989b, \apjs, 71, 433
\reference{} Pogge, R. W. 1992, in Astronomical CCD Observing and Reduction
             Techniques, ed. S. B. Howell (San Francisco: ASP), 195
\reference{} Pogge, R. W. \& Martini 2000, in preparation
\reference{} Ptak, A., Serlemitsos, P., Yaqoob, T., \& Mushotzky, R. F. 
             1999, \apjs, 120, 179
\reference{} Sankrit, R., Sembach, K.R.,  \& Canizares, C.R. 1999, 
             \apj, in press (astro-ph/9907406)
\reference{} Sargent, W. L. W., Young, P. J., Boksenberg, A., Shortridge, K., 
             Lynds, C. R., \& Hartwick, F. D. A. 1978, \apj, 221, 731
\reference{} Schmitt, H. R. \& Kinney, A.L. 1996, \apj, 463, 498
\reference{} Serlemitsos, P., Ptak, A., \& Yaqoob, T. 1996, in The Physics
             of LINERs in View of Recent Observations, ed. M. Eracleous,
             A. Koratkar, C. Leitherer, \& L. Ho (San Francisco: ASP), 70
\reference{} Shields, J. C. 1992, \apjl, 399, L27
\reference{} Stauffer, J. R. 1982, \apj, 262, 66
\reference{} Terashima, Y. 1999, Advances in Space Research, 23 (5-6), 851
\reference{} Terashima, Y., Ho, L. C., Ptak, A. F., Mushotzky, R. F., 
             Serlemitsos, P. J., Yaqoob, T., \& Kunieda, H. 1999, \apj, 
             submitted
\reference{} Terashima, Y., Kunieda, H., Misaki, K., Mushotzky, R. F., 
             Ptak, A., \& Reichert, G. A. 1998, \apj, 503, 212
\reference{} Terlevich, R. \& Melnick, J. 1985, \mnras, 213, 841
\reference{} Tsvetanov, Z., Morse, J. A., Wilson, A. S., \& Cecil, G. 1996, 
             \apj, 458, 172
\reference{} Tully, R. B. 1988, Nearby Galaxies Catalog 
             (Cambridge: Cambridge Univ. Press)
\reference{} van Dokkum, P. G. \& Franx, M. 1996, \aj, 110, 2027
\reference{} Watson, W. D. \& Wallin, B. K. 1994, \apjl, 432, L35
\reference{} Weaver, K. A., Wilson, A. S., Henkel, C., \& Braatz, J. A. 
             1999, \apj, 520, 130
\reference{} Wiklind, T. \& Henkel, H. 1990, \aap, 227, 394
\reference{} Wilkes, B. J., Schmidt, G. D., Smith, P. S., Mathur, S., \& 
             McLeod, K. K. 1995, \apjl, 455, L13
\reference{} Wilson, A. S. \& Tsvetanov, Z. I. 1994, \aj, 107, 1227
\reference{} Wilson, A. S., Braatz, J. A., Heckman, T. M., Krolik, J. H., \&
             Miley, G. K. 1993, \apjl, 419, L61
\reference{} Wrobel, J. M. 1984, \apj, 284, 531
\reference{} Zirbel, E. L. \& Baum, S. A. 1998, \apjs, 114, 177
\end{references}
\end{document}


\def\p{\phantom{0}}
\def\q{\phantom{-}\;\,}

\begin{deluxetable}{lllccclcr}
\tablenum{1}
\tablewidth{7.75in}
\tablecolumns{9}
\tablecaption{Sample Galaxies and Their Basic Properties}
\tablehead{
\colhead{} & 
\colhead{Hubble} & 
\colhead{$v_{\odot}$\tablenotemark{a}} & 
\colhead{Distance\tablenotemark{b}} &
\colhead{Scale}  &
\colhead{$B_T$\tablenotemark{a,c}} & 
\colhead{AGN} &
\colhead{$A_{\rm V}$\tablenotemark{e}} &
\colhead{Central} \nl
\colhead{Galaxy} & 
\colhead{Type\tablenotemark{a}} & 
\colhead{(km s$^{-1}$)} & 
\colhead{(Mpc)} &
\colhead{(pc/\arcsec)}  &
\colhead{(mag)} & 
\colhead{Class\tablenotemark{d}} &
\colhead{(mag)} &
\colhead{UV Flux\tablenotemark{f}} 
}
\startdata
NGC  404        & SA(s)0-:    &  $\p\p-46$ & \p2.4 & \p12  & 11.21 & 
L2\phantom{.0} & 0.28 &  $1.2\times 10^{-15}$ [1]  \nl 
NGC 1052        & E4          &  $\q1499$  &  17.8 & \p86  & 11.41 & L1.9
   & 0.16 &  $1.5\times 10^{-16}$ [2]  \nl 				 
NGC 3031  (M81) & SA(s)ab     &  $\p\p-37$ & \p3.6 & \p18  &\p7.89 & S1.5
   & 0.22 &  $1.5\times 10^{-15}$ [1]  \nl				 
NGC 3718        & SB(s)a pec  &  $\q\p993$ &  17.0 & \p82  & 11.59 & L1.9
   & 0.06 & $<7.0\times 10^{-18}$ [3]  \nl 				 
NGC 3998        & SA(r)0?     &  $\q1049$  &  21.6 &  105  & 11.61 & L1.9
   & 0.06 & $>9.8\times 10^{-15}$ [4]  \nl 				 
NGC 4036        & SA0-        &  $\q1396$  &  24.6 &  119  & 11.57 & L1.9
   & 0.10 & $<2.0\times 10^{-17}$ [5]  \nl 				 
NGC 4192  (M98) & SAB(s)ab    &  $\p-142$  &  16.8 & \p81  & 10.95 & L1.9
   & 0.14 & $<7.0\times 10^{-18}$ [3]  \nl 				 
NGC 4258 (M106) & SAB(s)bc    &  $\q\p450$ & \p7.2 & \p33  &\p9.10 & S1.9
   & 0.06 &  $7.0\times 10^{-17}$ [6]  \nl 
NGC 4374  (M84) & E1          &  $\q\p956$ &  16.8 & \p81  & 10.09 & 
L2\phantom{.0} & 0.14 & $<5.1\times 10^{-17}$ [7]  \nl 
NGC 4486  (M87) & E4+pec      &  $\q1282$  &  16.8 & \p81  &\p9.59 & 
L2\phantom{.0} & 0.13 &  $1.7\times 10^{-15}$ [8]  \nl 
NGC 4569  (M90) & SAB(rs)ab   &  $\p-235$  &  16.8 & \p81  & 10.26 & 
T2\phantom{.0} & 0.13 &  $1.0\times 10^{-14}$ [1]  \nl 
NGC 4579  (M58) & SAB(rs)b    &  $\q1521$  &  16.8 & \p81  & 10.48 & S1.9
   & 0.13 &  $3.3\times 10^{-16}$ [1]  \nl 
NGC 4594 (M104) & SA(s)a spin &  $\q1088$  &  20.0 & \p97  &\p8.98 & 
L2\phantom{.0} & 0.20 & $<1.2\times 10^{-16}$ [1]\tablenotemark{g}  \nl 
NGC 5005        & SAB(rs)bc   &  $\q\p948$ &  21.3 &  103  & 10.61 & L1.9
   & 0.06 & $<7.0\times 10^{-18}$ [3]  \nl

\tablenotetext{a\;}{The information was taken from the compilation of Ho et al. 
                    (1997a).}
\tablenotetext{b\;}{The distances, with three exceptions, were drawn from Tully 
                    (1988), who assumes $H_0=75~{\rm km~s^{-1}~Mpc^{-1}}$
                    (see \S5.2 of Ho et al. 1997a for further details). The
                    exceptions are NGC~3031 (Freedman et al. 1994) and
                    NGC~4258 (Herrnstein et al.  1999).}
\tablenotetext{c\;}{The total apparent blue magnitude of the galaxy.}
\tablenotetext{d\;}{Classification according to the ratios of the narrow 
                    emission lines and the presence of broad emission lines
                    (see Ho et al. 1997a, b for details). S = Seyfert, L =
                    LINER, T = transition object (intermediate between
                    LINER and H~II nucleus).  Objects with weak, broad
                    H$\alpha$ emission present are denoted as type ``1.9,''
                    while those without broad H$\alpha$ emission are
                    denoted as type ``2.''}
\tablenotetext{e\;}{Galactic visual extinction. Computed from the \ion{H}{1} 
                    column densities of Dickey \& Lockman (1990), assuming
                    $N_{\rm H}/A_{\rm V}=1.9\times 10^{21}~{\rm
                    cm^{-2}~mag^{-1}}$.}
\tablenotetext{f\;}{Measured UV flux density ($f_{\lambda}$), or upper limit, 
                    in units of ${\rm erg~cm^{-2}~s^{-1}~\AA^{-1}}$, at
                    2200\AA, except for NGC 3998, for which measurement is
                    at 1730\AA. Numbers in square brackets give the
                    reference, as follows: (1) Maoz et al. 1998; (2) from
                    archival {\it HST}/FOS spectra; (3) Maoz et al. 1995;
                    (4) Fabbiano et al. 1994; (5) Barth et al. 1998; (6) Ho
                    et al. 2000a; (7) Zirbel \& Baum 1998; (8) Maoz et
                    al. 1996.}
\tablenotetext{g\;}{See discussion in \S3.2 of the text.}
\enddata
\end{deluxetable}